\documentstyle[12pt,epsfig]{article}
\begin{document}

\begin{center}
{\large \bf FORM FACTORS OF DIMESOATOMS\\
FOR DISCRETE-CONTINUUM TRANSITIONS}
\vspace{.7cm}

S. Bakmaev and O. Voskresenskaya
\vspace{.05cm}

{\it Joint Institute for Nuclear Research,
141980 Dubna, Moscow Region, Russia\\}
\end{center}
\vspace{.7cm}

\begin{abstract}
{\small An approach for calculation of transition form factors of
hydrogen-like elementary atoms (EA) is proposed. A general formula
for bound-continuous transition form factors of EA is derived.
It is shown that these form factors can be represented in
the form of finite sum of terms with simple analytical structure and
may be numerically evaluated with arbitrary degree of accuracy.
As an application of the presented form factors, the calculation
of the spectra of products from EA ionization is considered.}
\end{abstract}
\vspace{.05cm}

\section{Introduction}

The problem of calculation of transition form factors from bound to
unbound states of hydrogen-like atoms [1] has a long history [2-4].
Nowadays, this problem became of great importance for the
interpretation of the data of DIRAC experiment at CERN which
aims to observe hydrogen-like EA\footnote{
Elementary atoms $A_{ab}$ are the Coulomb bound states of two
elementary particles. One can enumerate here
$A_{2e}$, $A_{e\mu}$, $A_{2\mu}$, $A_{e\pi}$, $A_{\mu\pi}$,
$A_{2\pi}$, $A_{\pi K}$, $A_{K K}$.} [5]
consisting of $\pi^\pm$ and/or $\pi^\mp$/$K^\mp$ mesons (dimesoatoms)
and measure the lifetime of $\pi^+\pi^-$ atoms ($A_{2\pi}$) with
accurace of 10\% [6-8].

The usual approach to this problem is based on the decomposition of
continuum wave functions into infinite series of partial waves and
calculation of transition form factors from initial bound state to
final continuum state with definite value of angular momentum [4].

In this approach only finite number of terms of infinite series are
taken into account in actual calculation leaving unsolved the problem of
estimation for contribution of omitted tail with infinite number of terms.

We would like to show in this paper that above mentioned transition
(bound to continuum) form factors of hydrogen-like EA may be
explicitly calculated without decomposition of final state into
infinite series of partial waves.

The plan of the paper is as follows. First of all in Section 2
we shall prove this statement for the simplest case when orbital
momentum of bound state is equal zero, i.e. we restrict of ourselves
with initial $nS$-states.  The generalization of this consideration
for the case of arbitrary initial states will be done in Section 3.
Finally in Section 4 we compute the spectra of $\pi^+\pi^-$ pairs
from ionization (dissociation) of $A_{2\pi}^{nS}$
as an application of the presented form factors.

\section{Transitions from $nS$-states}

We define the transition form factors with the help of the following
equation:  \begin{equation} S_{fi}(\vec q)=\int \psi_f^{\ast}(\vec
r)e^{i\vec q\vec r} \psi_i(\vec r)d^3r \label{e1}\, , \end{equation}
where $\psi_{i(f)}$ are the wave functions of initial (final) states
of hydrogen-like atoms, $\vec q$ is the transferred momentum.

For the case when  $i=n00\equiv nS$
\begin{eqnarray}
\psi_i=\psi_{n00}(\vec r)&=&\left(\frac{\omega^3}{\pi}\right)^{\frac{1}{2}}
\exp(-\omega r)\cdot\Phi(-n+1;2;2\omega r)\nonumber\\
&\equiv&\left(\frac{\omega^3}{\pi}\right)^{\frac{1}{2}}n^{-1}
\exp(-\omega r)\cdot L^{1}_{n-1}(2\omega r)
\label{e2}\, ,
\end{eqnarray}
where $\omega=\mu\alpha/n$; $\mu$ is the reduced mass and
$\alpha=1/137$ is the fine structure constant; $\Phi$ is the confluent
hypergeometric function and $L^{\lambda}_k$ are the associated Laguerre
polynomials.

The wave function of the final (continuum) state must be choose in the form
\cite{land}
\begin{eqnarray}
\psi_f(\vec r)=\psi_{\vec p}^{(-)}&=&c^{(-)}\exp(i\vec p\vec r)
\cdot\Phi\left[-i\xi,1,-i(pr+\vec p\vec r)\right]\,,
\label{e3}
\end{eqnarray}
$$c^{(-)}=(2\pi)^{-\frac{3}{2}}\exp\left(\frac{\pi\xi}{2}\right)
\Gamma(1+i\xi)\,,$$
$$\xi=\frac{\mu\alpha}{p}\,.$$

Now we use recurrence relation for the Laguerre polynomials \cite{ryzh}
\begin{equation}
L_k^{\lambda+1}(x)=\frac{1}{x}\left[(k+\lambda+1)L_{k-1}^{\lambda}(x)
-(k+1)L_{k}^{\lambda}(x)\right]
\label{e4}
\end{equation}
and their representation with the help of the generating function
\begin{equation}
L_k^{\lambda}(x)=\Delta_z^{(k)}\left[(1-z)^{-(\lambda+1)}\exp
\left(\frac{xz}{z-1}\right)\right]
\label{e5}\, ,
\end{equation}
where operator $\Delta_z^{(k)}$ is defined as follows:
\begin{equation}
\Delta_z^{(k)}\left[f(z)\right]=\frac{1}{k!}
\left.\left(\frac{d^k}{dz^k}f(z)\right)\right\vert_{z=0}
\label{e6}\, .
\end{equation}

Then
\begin{equation}
\psi_i(r)=\frac{1}{2r}\left(\frac{\omega}{\pi}\right)^{\frac{1}{2}}
\left[\Delta_z^{(n-1)}-\Delta_z^{(n)}\right]
\left[(1-z)^{-1}\exp\left[-\omega(z)r\right]\right]
\label{e7}\, ,
\end{equation}
\begin{equation}
\omega(z)=\omega\cdot \frac{1+z}{1-z}\,.
\label{e11}
\end{equation}

The substitution of Eqs. (\ref{e3}) and (\ref{e7}) to (\ref{e1}) gives
\begin{equation}
S_{\vec p,noo}(\vec q)=\frac{1}{2}\left(\frac{\omega}{\pi}\right)^{\frac{1}{2}}
c^{(-)}
\left[\Delta_z^{(n-1)}-\Delta_z^{(n)}\right]
\left[\frac{J(\vec q,\vec p,z)}{(1-z)}\right]
\label{e8}\, ,
\end{equation}
\begin{equation}
J(\vec q,\vec p,z)=\int \frac{d^3r}{r}
\Phi\left[i\xi,1,i(pr+\vec p\vec r)\right]
\exp[i(\vec q-\vec p)\vec r-\omega(z) r]\,.
\label{e9}
\end{equation}

The last integral is easily calculated  using integral representation
for the hypergeometrical functions (see e.g. \cite{nord}).

The result reads
\begin{equation}
J(\vec q,\vec p,z)=4\pi[\omega^2(z)+\vec\Delta^2]^{-1+i\xi}
\left[[\omega(z)-ip]^2+q^2\right]^{-i\xi}\,,
\label{e10}
\end{equation}
where $\vec\Delta=\vec q-\vec p$.

Taking into account the definition (\ref{e11})
and obvious the relation
\begin{equation}
\Delta_z^{(n)}\left[zf(z)\right]=\Delta_z^{(n-1)}f(z)\,,
\label{e12}
\end{equation}
(\ref{e8}) may be rewritten in the form
\begin{equation}
S_{\vec p,noo}(\vec q)=-4(\pi\cdot\omega)^{\frac{1}{2}}c^{(-)}
\left(\Delta_z^{(n)}-2\Delta_z^{(n-1)}+\Delta_z^{(n-2)}\right)
\left(D_1^{-1+i\xi}D_2^{-i\xi}\right)
\label{e13}\, ,
\end{equation}
$$D_1=(1+z^2)(\omega^2+\vec\Delta^2)-2z(\vec\Delta^2-\omega^2)\,,$$
$$D_2=(\omega-ip)^2+q^2-2z(q^2-p^2-\omega^2)+
z^2\left((\omega+ip)^2+q^2\right)\,.$$

Using the definition of the Gegenbauer polynomials \cite{ryzh,abram}
\begin{equation}
(1-2xz+z)^{-\lambda}=\sum_{k=0}^{\infty}C_k^{(\lambda)}(x)\cdot z^k\,,
\label{e14}
\end{equation}
it is easy to obtain
\begin{equation}
D_1^{-1+i\xi}=(\Delta^2+\omega^2)^{-1+i\xi}
\sum_{k=0}^{\infty}C_k^{(1-i\xi)}(u)\cdot z^k
\label{e15}\,,
\end{equation}
$$u=\frac{\Delta^2-\omega^2}{\Delta^2+\omega^2}\,;$$
\begin{equation}
D_2^{-i\xi}=[(\omega-ip)^2+q^2]^{-i\xi}
\sum_{k=0}^{\infty}C_k^{(i\xi)}(v)\cdot w^k\cdot z^k
\label{e16}\,,
\end{equation}
$$v=\frac{q^2-p^2-\omega^2}{\sqrt{[(\omega-ip)^2+q^2][(\omega+ip)^2+q^2]}}\,,$$
\begin{equation}
w=\sqrt{\frac{(\omega+ip)^2+q^2}{(\omega-ip)^2+q^2}}\,.
\end{equation}

At least, with the help of the relations
\begin{equation}
\Delta_z^{(n)}[f_1(z)f_2(z)]
=\sum_{k=0}^{n}
\left[\Delta_z^{(n-k)}f_1(z)\right]\left[\Delta_z^{(k)}f_2(z)\right]
\label{e17}
\end{equation}
and
\begin{equation}
C_n^{(\lambda)}(x)-C_{n-1}^{(\lambda)}(x)=\frac
{\Gamma(\lambda-\frac{1}{2})\Gamma(n-1+2\lambda)}
{\Gamma(2\lambda-1)\Gamma(n+\lambda-\frac{1}{2})}
\cdot P_n^{(\lambda-\frac{3}{2},\lambda-\frac{1}{2})}(x)\,,
\label{e18}
\end{equation}
where $P_n^{(\lambda-\frac{3}{2},\lambda-\frac{1}{2})}(x)$ are the Jacobi
polynomials, we finally obtain
\vspace*{.25cm}
\begin{equation}
S_{\vec p,noo}(\vec q)=-2(\pi\cdot\omega)^{\frac{1}{2}}\cdot c^{(-)}
\cdot(\omega^2+\Delta^2)^{-1+i\xi}
\label{e19}\end{equation}
$$\times[(\omega-ip)^2+q^2]^{-i\xi}
\frac{\Gamma\left(\frac{1}{2}- i\xi\right)}
{\Gamma\left(1- 2i\xi\right)}
\sum_{k=0}^{n}w^kC_k^{(i\xi)}(v)$$

$$
\times\left[\frac{\Gamma\left(n-k+1-2i\xi\right)}
{\Gamma\left(n-k+\frac{1}{2}- i\xi\right)}
P_{n-k}^{(-\frac{1}{2}-i\xi,\frac{1}{2}-i\xi)}(u)
-\frac{\Gamma\left(n-k- 2i\xi\right)}
{\Gamma\left(n-k-\frac{1}{2}- i\xi\right)}
P_{n-k-1}^{(-\frac{1}{2}-i\xi,\frac{1}{2}-i\xi)}(u)\right].
$$

Thus, the result of the calculation for transition form factors of
hydrogen-like atoms for the case  $nS$-continuum transitions is
expressed in terms of the classical polynomials and may be easily
evaluated numerically with arbitrary degree of accuracy \cite{voskr1}.

\section{General case}

In the previous Section it have been shown that the form factors of
transitions from $nS\; (n00)$ - states of hydrogen-like atoms
to the state of continuous spectra with definite value
of relative momenta $\vec p$ may be expressed in the terms of the
classical polynomials in a rather simple way. Below this result is
generalized for the case of  transition from arbitrary initial bound
states.

The transition form factors are defined as follows:
\begin{equation}
S_{fi}(\vec q)=\int \psi_f^{\ast}(\vec r)e^{i\vec q\vec r}
\psi_i(\vec r)d^3r
\label{e1}\, ,
\end{equation}
Here, $\psi_{i(f)}$ are the wave functions of initial (final) states.

According to \cite{sommer} (see also \cite{land}), the final state wave
function must be choose in the form
\begin{equation}
\psi_i(\vec r)\equiv\psi_{n00}(\vec r)
\label{e2}\, .
\end{equation}
For arbitrary initial bound state
\begin{equation}
\psi_i(\vec r)=\psi_{nlm}(\vec r)=
Y_{lm}\left(\frac{\vec r}{r}\right)R_{nl}(r)\,,
\label{351}
\end{equation}
\begin{eqnarray}
R_{nl}(r)&=&\frac{2\omega^{\frac{3}{2}}}{\Gamma(2l+2)}
\left[\frac{\Gamma(n+l+1)}{n\Gamma(n-l)}\right]^{\frac{1}{2}}
\cdot(2\omega r)^l
\cdot\Phi(-n+l+1,2l+2;2\omega r)\cdot \exp(-\omega r)\nonumber
\end{eqnarray}
$$~~~~=2\omega^{\frac{3}{2}}
\left[\frac{\Gamma(n-l)}{n\Gamma(n+l+1)}\right]^{\frac{1}{2}}
\cdot(2\omega r)^l
\cdot L^{2l+1}_{n-l-1}(2\omega r)\cdot \exp(-\omega r)\,,$$
\begin{equation}
\omega=\frac{\mu\alpha}{n}\,,
\end{equation}
where $L^{2l+1}_{n-l-1}$ are the associated Laguerre polynomials.

Making use of the recurrence relations \cite{ryzh,abram}
\begin{equation}
L_k^{\lambda+1}(x)=\frac{1}{x}\left[(k+\lambda+1)L_{k-1}^{\lambda}(x)
-(k+1)L_{k}^{\lambda}(x)\right]
\label{e4}
\end{equation}
and the representation of the Laguerre polynomials
in terms of the generating function
\begin{equation}
L_k^{\lambda}(x)=\Delta_z^{(k)}\left[(1-z)^{-(\lambda+1)}\exp
\left(\frac{xz}{z-1}\right)\right]
\label{e5}\, ,
\end{equation}
where operator $\Delta_z^{(k)}$ is defined as follows
\begin{equation}
\Delta_z^{(k)}\left[f(z)\right]=\frac{1}{k!}
\left.\left(\frac{d^k}{dz^k}f(z)\right)\right\vert_{z=0}
\label{e6}\, ,
\end{equation}
let us rewrite the radial part of initial state wave
function in the form
\begin{equation}
R_{nl}=\frac{\omega^{\frac{1}{2}}}{r}
\left[\frac{\Gamma(n-l)}{n\Gamma(n+l+1)}\right]^{\frac{1}{2}}
\cdot(2\omega r)^l
\cdot\left[(n+l)\Delta_z^{(n-l-2)}-(n-l)\Delta_z^{(n-l-1)}\right]
\label{e7}
\end{equation}
$$\times\left[(1-z)^{-(l+1)}\exp\left(-\omega(z)r\right)\right]\,,$$
\begin{equation}
\omega(z)=\omega\cdot(1+z)(1-z)^{-1}
\label{e11}
\end{equation}
more convenient for the further calculations.

Then it is not difficult to see that transition form factors
(\ref{e1}) may be represent as a linear combination of the quantities
\begin{equation}
I_{lm}^{j}=\Delta_z^{(j)}\left[(1-z)^{-(2l+1)}
J_{lm}(\vec q,\vec p,z)\right]
\label{eq:352}\, ,
\end{equation}
\begin{eqnarray}
J_{lm}(\vec q,\vec p,z)&=&\int \frac{d^3r}{r}Y_{lm}
\left(\frac{\vec r}{r}\right)
\Phi\left[i\xi,1;i(pr+\vec p\vec r)\right]
\label{eq:353}
\end{eqnarray}
$$\times\exp[i(\vec q-\vec p)\vec r-\omega(z)r](2\omega r)^l
\exp\left[-\omega(z)r\right]\,,$$
$$j=n-l-2,\;n-l-1\,.$$

In order to calculate (\ref{eq:353}), it is useful to represent the
hypergeometrical function in (\ref{351}) in the form
\begin{eqnarray}
\Phi\left[i\xi,1;i(pr+\vec p\vec r)\right]&=&-\frac{1}{2\pi i}
\oint\limits_{C}^{} (-t)^{i\xi-1}(1-t)^{-i\xi}
\cdot\exp[i\cdot t(pr+\vec p\vec r)]dt\,.
\label{eq:354}
\end{eqnarray}

Using the following relations
\begin{equation}
\label{eq:355}
\exp(i\vec\tau\vec r)=4\pi\sum_{l=0}^{\infty}\sum_{m=-l}^{l}
i^{l}Y_{lm}\left(\frac{\vec \tau}{\tau}\right)
Y^{\ast}_{lm}\left(\frac{\vec r}{r}\right)j_{l}(\tau r)\,,
\end{equation}
\begin{equation}
\label{eq:356}
j_{l}(x)=\sqrt{\frac{\pi}{2x}}J_{l+\frac{1}{2}}(x)\,,
\end{equation}
\begin{equation}
\label{eq:357}
\int\limits_{0}^{\infty}r^{l+\frac{1}{2}}J_{l+\frac{1}{2}}(\tau r)
e^{-\bar\omega\cdot r}dr=
\frac{(2\tau)^{l+\frac{1}{2}}\Gamma(l+1)}
{\sqrt{\pi}(\tau^2+\bar \omega^2)^{l+1}}\,,
\end{equation}
where
\begin{equation}
\label{eq:358}
\vec\tau=\vec q-\vec p(1-t),\quad \bar\omega=\omega(z)-ip\cdot t\,,
\end{equation}
after simple calculations we find
\begin{eqnarray}
J_{lm}(\vec q,\vec p,z)&=&-\frac{\Gamma(l+1)}{2\pi i}
\oint\limits_{C}^{ }dt(-t)^{i\xi-1}(1-t)^{-i\xi}
\cdot\frac{4\pi(4i\omega)^lY_{lm}\left(\vec \tau/\tau\right)\tau^l}
{(\tau^2+\bar \omega^2)^{l+1}}\,.
\label{eq:359}
\end{eqnarray}

It is easy to check that
$$\tau^2+\bar \omega^2=a(1-t)+c\cdot t\,,$$
\begin{equation}
\label{eq:362}
a=\omega^2(z)+\vec\Delta^2\,,\quad c=\left[\omega(z)-ip\right]^2+q^2\,.
\end{equation}

Further, according to \cite{war}, we get
\begin{eqnarray}
\label{eq:360}
Y_{lm}\left(\frac{\vec \tau}{\tau}\right)\tau^l
=\sum_{l_1=0}^{l}q^{l_1}(-p)^{l-l_1}
\end{eqnarray}
$$\times\left[\frac{4\pi\Gamma(2l+2)}
{\Gamma(2l_1+2)\Gamma(2l-2l_1+2)}\right]^{\frac{1}{2}}
(1-t)^{l-l_1}\left[Y_{l_1}\left(\frac{\vec q}{q}\right)\otimes
Y_{l-l_1}\left(\frac{\vec p}{p}\right)\right]_{lm}\,,$$
\begin{eqnarray}
\label{eq:361}
\left[Y_{l_1}\left(\frac{\vec q}{q}\right)\otimes
Y_{l-l_1}\left(\frac{\vec p}{p}\right)\right]_{lm}
=\sum_{m_1+m_2=m}C^{lm}_{l_1m_1(l-l_1)m_2}
\cdot Y_{l_1m_1}\left(\frac{\vec q}{q}\right)\cdot
Y_{(l-l_1)m_2}\left(\frac{\vec p}{p}\right)\,.
\end{eqnarray}
Taking into account (\ref{eq:362}) and (\ref{eq:360}), it is easy
to see that (\ref{eq:359}) is the superposition of the quantities
\begin{equation}
-\frac{1}{2\pi i}\oint\limits_{C}^{} \frac{t^{i\xi-1}(1-t)^{-i\xi+l-l_1}}
{[a(1-t)+ct]^{l+1}}=
\end{equation}
\begin{eqnarray}
&&=a^{-(l+1)}\frac{\Gamma(1-i\xi+l-l_1)}{\Gamma(1-i\xi)}
F\left(i\xi,l+1;l-l_1+1;1-c/a\right)\nonumber\\
&&=a^{i\xi-l-1}c^{-i\xi}\frac{\Gamma(1-i\xi+l-l_1)}
{\Gamma(1-i\xi)}
F\left(i\xi,-l_1;l-l_1+1;1-a/c\right)\nonumber\\
&&=a^{i\xi-l-1}c^{-i\xi}\frac{\Gamma(l-l_1+1)\Gamma(l+1-i\xi)}
{\Gamma(l+1)\Gamma(1-i\xi)}F(i\xi,-l_1;i\xi-l;a/c)\nonumber\\
&&=\sum_{s=0}^{l_1}(-1)^{l-s}
\frac{\Gamma(i\xi+s)\Gamma(l_1+1)}
{\Gamma(l_1-s+1)\Gamma(i\xi-l+s)\Gamma(s+1)\Gamma(l+1)}
a^{i\xi+s-l-1}c^{-s-i\xi}\nonumber\\
&&=(1-z)^{2l+2}\sum_{s=0}^{l_1}(-1)^{l-s}
\frac{\Gamma(i\xi+s)\Gamma(l_1+1)}
{\Gamma(l_1-s+1)\Gamma(i\xi-l+s)\Gamma(s+1)\Gamma(l+1)}\nonumber
\label{eq:363}
\end{eqnarray}
$$\times D_1^{i\xi+s-l-1}D_2^{-s-i\xi}\,,$$
where $D_{1,2}$ are defined as follows:
\begin{equation}
D_1=(1+z^2)(\omega^2+\vec\Delta^2)-2z(\vec\Delta^2-\omega^2)\,,
\label{e3}
\end{equation}
$$D_2=(\omega-ip)^2+q^2-2z(q^2-p^2-\omega^2)+z^2[(\omega+ip)^2+q^2]\,.$$

The further calculations are the same as in the Section 2.

Omitting the simple but cumbersome algebra, let us present the final
expression for transition form factors:

\begin{eqnarray}
S_{\vec p,nlm}(\vec q)&=&4\pi\cdot 2^{2l}i^l\omega^{l+\frac{1}{2}}
\left[\frac{\Gamma(n-l)}{n\Gamma(n+l+1)}\right]^{\frac{1}{2}}\nonumber\\
\label{eq:364}
\end{eqnarray}
$$
\times\sum_{s=0}^{l}G_{lms}(\vec p,\vec q)H_{nls}(\vec p,\vec q)
(\omega^2+\Delta ^2)^{i\xi+s-l-1}[(\omega-ip)^2+q^2]^{-s-i\xi}\,;
$$
\begin{eqnarray}
\label{eq:365}
G_{lms}(\vec p,\vec q)&=&(-1)^{l-s}
\frac{\Gamma(i\xi+s)}{\Gamma(i\xi-l+s)\Gamma(s+1)}
\end{eqnarray}
$$\times\sum_{l_1=s}^{l}
\left[\frac{4\pi\Gamma(2l+2)}{\Gamma(2l_1+2)\Gamma(2l-2l_1+2)}\right]^
{\frac{1}{2}}
\frac{\Gamma(l_1+1)}{\Gamma(l_1-s+1)}q^{l_1}(-p)^{l-l_1}$$
$$\times\left[Y_{l_1}\left(\frac{\vec q}{q}\right)\otimes
Y_{l-l_1}\left(\frac{\vec p}{p}\right)\right]_{lm}\,;$$
\begin{eqnarray}
\label{eq:366}
H_{nls}(\vec p,\vec q)&=&(n+l)F_{n_1ls}(\vec p,\vec q)-
(n-l)F_{n_2ls}(\vec p,\vec q)\,;
\end{eqnarray}
$$n_1=n-l-1\,,\quad n_2=n-l-2\,;$$
\begin{eqnarray}
\label{eq:367}
F_{n_{1(2)}ls}(\vec p,\vec q)&=&
\frac{\Gamma(l-s+\frac{1}{2}-i\xi)}{\Gamma(2l-2s+1-2i\xi)}
\sum_{k=0}^{n_{1(2)}}w^kC_k^{(i\xi+s)}(v)
\end{eqnarray}
$$\times\frac{\Gamma(n_{1(2)}-k+2l-2s+1-2i\xi)}
{\Gamma(n_{1(2)}-k+l-s+\frac{1}{2}-i\xi)}
 P_{n_{1(2)}-k}^{(l-s-\frac{1}{2}-i\xi,l-s+\frac{1}{2}-i\xi)}(u)\,.$$\\

Thus, the form factors for transition from arbitrary bound states of
hydrogen-like EA to the ``$\vec p-state$'' of continuous spectra
are represented as the superposition of finite number of terms with
simple analytical structure and can be also calculated with
arbitrary degree of accuracy.

Eqs. (\ref{eq:364})-(\ref{eq:367}) are the  basic  results of this
study. They are the generalization of the results of Section 2 and
Refs. [3].

\section{Applications}

The mentioned above results are necessary for the calculations of
the spectra of products from $A_{\pi^+\pi^-}$/$A_{\pi^\pm K^{\mp}}$
ionization which is/will exploited to observe
$A_{\pi^+\pi^-}$/$A_{\pi^\pm K^{\mp}}$ atoms and to measure its
ground state lifetime [6,9].

These  spectra may be  represented  and  computed  as
\begin{equation}
\frac{d^3\sigma_{(nlm\to\vec p)}}{d^3\vec  p}=
\frac{1}{(2\pi)^3}\int d^2 q\, \sigma_0(\vec q)
\left [S_{\vec p,nlm}(\vec  p,\vec  q_1 )-
S_{\vec p,nlm}(\vec  p,\vec  q_2 )\right ],
\end{equation}
\begin{equation}
\vec q_1=\frac{\vec  qm_1}{M},\quad \vec q_2=-\frac{\vec  qm_2}{M},\quad  M=m_1+m_2,
\end{equation}
\begin{equation}
\sigma_0(\vec q)=\left(\frac{2\alpha}{q^2}\right)^2\left\{ [Z-F_{el}(\vec q)]^2+
ZF_{inel}(\vec q)\right\},
\end{equation}
$$F_{el}(\vec q=0)=Z.$$
Here, $\sigma_{(nlm\to\vec p)}$ is the cross section
of dimesoatoms for transitions from $|nlm\rangle$ states to
continuum; $S_{\vec p,nlm}$ is the corresponding transition form factor;
$m_{1,2}$ are the masses of the $\pi^\pm$ and
$\pi^\mp/K^\mp$ mesons; $\alpha$ is the fine  structure   constant;
$F_{el}(\vec q)$ and $F_{inel}(\vec q)$ are the elastic and
inelastic atomic form factors of the target atom respectively,
$Z$ is its atomic number.

\begin{figure}[p]

\begin{center}

\epsfig{file=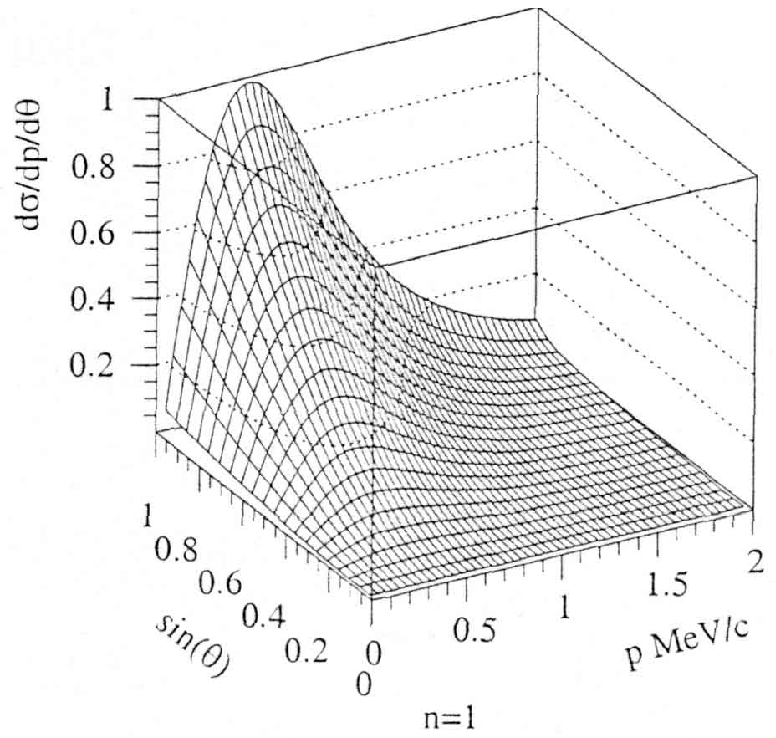}

\caption{Spectrum of pions from ionization of
$A_{2\pi}^{1S}$ atoms.}
\label{void}

\end{center}

\end{figure}

The results of such calculations are illustrated by Figures 1,2.
In these Figures  momentum ($p$) and angular ($\theta$) distributions
\begin{equation}
\frac{d\sigma_{(nlm\to\vec p)}}{d\theta  dp}=
2\pi \sin\theta \, p^2
\frac{d^3\sigma_{(nlm\to\vec p)}}{d^3\vec  p}
\end{equation}
in $\pi^+\pi^-$ pairs from dissociation of $\pi^+\pi^-$ atoms
in the Coulomb field of target atoms
are shown for the initial states of $A_{2\pi}$ with principal quantum
numbers $n=1$ (Figure~\ref{void}) and $n=10$ (Figure~2).

\begin{figure}[p]

\begin{center}

\epsfig{file=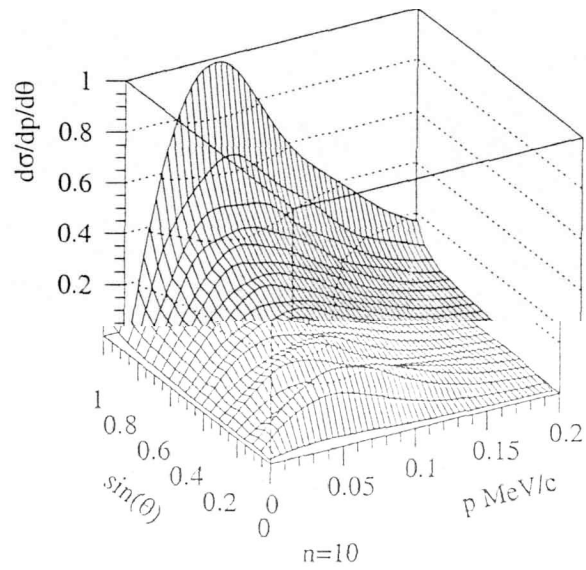}

\caption{Spectrum of pions from ionization of
$A_{2\pi}^{10S}$ atoms.}
\label{void1}

\end{center}

\end{figure}

\clearpage
\section*{Acknowledgments}
\bigskip

The authors are grateful to Alexander Tarasov and Leonid Afanasyev
for useful discussions.

\end{document}